\documentclass[prb,preprintnumbers,amsmath,amssymb,floatfix,aps,twocolumn]{revtex4}
\usepackage{graphicx}

\usepackage{hhline}
\usepackage{hyperref}
\usepackage{amsmath}
\usepackage{amsbsy}
\usepackage{gensymb}
\usepackage{float}

\begin{document}

\title{Highly anisotropic thermoelectric properties of carbon sulfide monolayers}

\date{\today} 

\author{J. W. Gonz\'alez\footnote{Corresponding author:sgkgosaj@ehu.eus }}
\affiliation{
Centro de F\'{i}sica de Materiales (CSIC-UPV/EHU)-Material Physics Center (MPC), Donostia International 
Physics Center (DIPC), Departamento de F\'{i}sica de Materiales, Fac. Qu\'{i}micas UPV/EHU. Paseo Manuel de Lardizabal 5, 20018, San Sebasti\'an-Spain.
}

\begin{abstract}
Strain engineering applied to carbon monosulphide monolayers allows 
to control the bandgap, controlling electronic and thermoelectric 
responses. 
Herein, we study the semiconductor-metal phase transition of this 
layered material driven by strain control on the basis of 
first-principles calculations. 
We consider uniaxial and biaxial tensile strain and we
find a highly anisotropic electronic and thermoelectonic 
responses depending on the direction of the applied strain.
Our results indicate that strain-induced response could be an 
effective method to control the electronic response and the 
thermoelectric performance.
\end{abstract}

\maketitle

\section{\label{sec:intro} Introduction}

In recent years new phenomena emerging from two-dimensional materials 
have been explored. 
Despite the exciting new electronic properties 
that make it possible to design new 
applications\cite{jariwala2014emerging,fiori2014electronics}, 
one of the biggest challenges ahead is to control bandgap in a systematic 
way. Several solutions are currently being considered such as 
doping\cite{mahmood2015nitrogenated,gonzalez2015electron}, 
stacking orders\cite{dai2014bilayer,cortes2018stacking,chico2017spin,gonzalez2010electronic} 
or strain\cite{frisenda2017biaxial,zhang2014phosphorene}.
Research into two-dimensional (2D) materials goes beyond
graphene\cite{bhimanapati2015recent,das2015beyond}. On one hand, elements 
of the same carbon group have been considered to produce layered 
materials, for instance the silicene, germanene or stanene 
monolayers\cite{balendhran2015elemental,lalmi2010epitaxial}. 
On the other hand, the quest for new electronic properties
in layered materials  has been extended to the various 
phosphorus-based assembles, in particular the black phosphorus monolayers
or phosphorene may impact on future technologies\cite{li2014black}. 
Following these trends, our goal is to demonstrate the interesting 
properties of the recently predicted carbon sulfide monolayer 
(CS monolayer) which is isoelectronic to phosphorene\cite{alonso2017stable} 
and its possible control under external strain.

In two-dimensional systems, strain can be applied indirectly by using thermal 
variations of the substrate\cite{frisenda2017biaxial}  or directly
by mechanical 
deformations\cite{lee2012optical,caneva2018mechanically}.
Theoretical studies predict that most 2D materials can tolerate 
strain values above 10\% without 
ruptures\cite{frisenda2017biaxial,bertolazzi2011stretching,caneva2018mechanically}.
For example, the graphene monolayers can 
easily tolerate strain above 25\%\cite{lee2008measurement}.
Our aim is to provide a general picture of the connection between 
electronic and thermoelectronic properties upon external strain,
highlighting the efficiency of the strain applied in certain directions 
to control the bandgap. 

Herein, we employ first-principles calculations to study the electronic 
and thermoelectronic response upon strain for the carbon 
sulfide monolayers.
Calculations show that the CS monolayer is a semiconductor stable 
at room temperature with an indirect bandgap\cite{alonso2017stable}. Its 
structure is composed by single layer with two-dimensional honeycomb 
puckered structure where each atom is bonded to three neighbors. 
Because the CS monolayer has a structure similar to phosphorene, 
therefore we anticipate a similar improvement in mechanical 
flexibility\cite{wei2014superior} and highly anisotropic 
electronic properties\cite{li2014black}.

\begin{figure}[b!]
\centering
\includegraphics[clip,width=0.47\textwidth,angle=0]{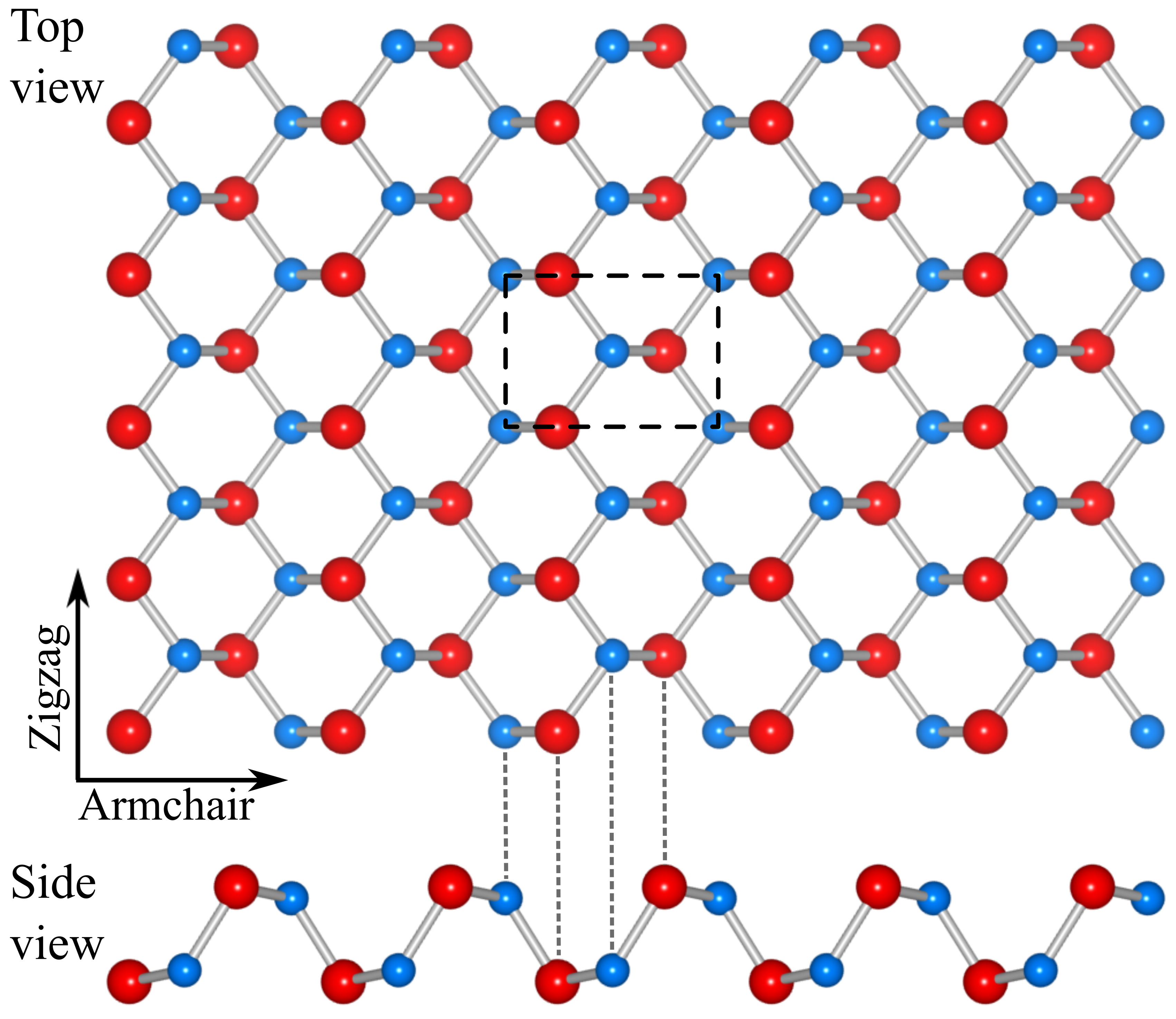} 
\caption{(Color on-line) Ball and stick model for the atomic structure of 
CS monolayer sheet. The unit cell is indicated with a dashed line. 
The armchair edge is parallel to the x-axis and the zigzag edge 
is parallel to the y-axis.}
\label{Fig:scheme}
\end{figure}

\section{Methodology}
Density functional theory calculations were performed using the 
plane-wave self-consistent field plane-wave implemented in the 
{\sc Quantum ESPRESSO} package\cite{giannozzi2009quantum} with 
the generalized gradient approximation of Perdew-Burke-Ernzerhof (PBE)
exchange-correlation functional\cite{perdew1996generalized,pseudo}. 
Self-consistent charge calculations are converged up to a tolerance 
of $10^{-8}$ for the unit cell. Monolayers are repeated periodically
separated by 20~\AA{} of empty space in the perpendicular direction. 
A fine k-grid of $20\times 20 \times 1$ Monkhorst-Pack is used to sample the 
Brillouin zone and the orbitals were expanded in plane waves until 
a kinetic energy cutoff of $680$ eV. All atoms are allowed to relax 
within the conjugate gradient method until forces have been converged 
with a tolerance of $10^{-3}$ eV/\AA{}. 
For the calculation of the transport coefficients a denser k-grid of 
$50\times50\times1$ Monkhorst-Pack is used to sample the Brillouin zone.

In general, an accurate description of the bandgap requires 
sophisticated  semi-local exchange-correlation approaches as the 
GW approximation\cite{tran2009accurate,tran2014layer}.
However, under experimental conditions, samples are subject to external factors such as doping or interaction with substrates. Therefore, 
the electronic screening is significantly weaker than in the isolated case 
and as a consequence the bandgap tends to  be 
smaller\cite{hybertsen1986electron}. 
In order to consider the external factors, we calculate the 
thermoelectric coefficients using PBE-DFT band structure, that 
can be used as a lower limit of the bandgap\cite{fei2014enhanced}.

The electronic transport coefficients are derived from the 
electronic band structure based on the semiclassical Boltzmann 
transport theory within the constant relaxation time approximation 
(RTA), as implemented in the {\sc BoltzTraP} 
code\cite{madsen2006boltztrap}. In this approximation, the 
relaxation time is a constant and therefore the thermopower or 
Seebeck coefficient $S$ is independent of the relaxation time.
The constant relaxation time approximation has been successfully 
used to describe the transport coefficients of a wide range of 
thermoelectric 
materials \cite{li2016anisotropic,hung2015diameter,yang2014relationship,peng2011electronic}.

The Seebeck coefficient at the temperature $T$ and 
chemical potential $\mu$ can be expressed
as\cite{madsen2006boltztrap,scheidemantel2003transport},
\begin{equation}
    S=\frac{q k_B}{\sigma} \int d\varepsilon 
    \left( -\frac{\partial f_0}{\partial \varepsilon} \right)
    \Xi \left( \varepsilon\right) 
    \frac{ \varepsilon - \mu }{k_B T},
\end{equation}
being the scalar conductivity $\sigma$ defined by
\begin{equation}
    \sigma = q^2
    \int d\varepsilon 
    \left( -\frac{\partial f_0}{\partial \varepsilon} \right)
    \Xi \left( \varepsilon\right),
\end{equation}
with the transport distribution $\Xi$ defined as
\begin{equation}
    \Xi = \sum _{\Vec{k}} \Vec{v}_{\Vec{k}}\Vec{v}_{\Vec{k}} 
    \tau_{\Vec{k}},
\end{equation}
where $q$ is the carrier charge, $f_0$ the Fermi distribution,
$k_B$ the Boltzmann constant,
$\tau_{\Vec{k}}=\tau_0$ the constant relaxation time
and $\Vec{v}_{\Vec{k}} = \frac{1}{\hbar} 
\frac{\partial \varepsilon_{\Vec{k}}}{\partial \Vec{k}}$ 
the group velocity  of the state labeled with $\Vec{k}$.

\section{Results and Discussion}
Figure \ref{Fig:scheme} shows the most stable CS monolayer structure, 
corresponding to a hexagonal structure with puckered sheets of bounded 
atoms in the same way as phosporene.  
Similar structures can be found in group IV monochalcogenides 
(for instance GeSe, GeS, SnSe,
SnS)\cite{singh2014computational,hu2015gese,vaughn2010single}, and 
group V semiconductors (PN or
AsN)\cite{zhu2015designing,zhang2016tinselenidene,xiao2016prediction}.
Each sulfur atom on the CS monolayer is bonded to three carbon atoms and 
vice versa\cite{alonso2017stable}. An out-plane bond of length 1.84~\AA{} and 
two in-plane bonds of length 1.75~\AA{} are observed. The angles 
between the bonds are 102$\degree$ and 105$\degree$. 
With the optimized parameters, the relaxed in-plane lattice vectors 
are $a=4.03$~\AA{} and $b=2.77$~\AA{}. 

\begin{figure}[!ht]
\centering
\includegraphics[clip,width=0.47\textwidth,angle=0]{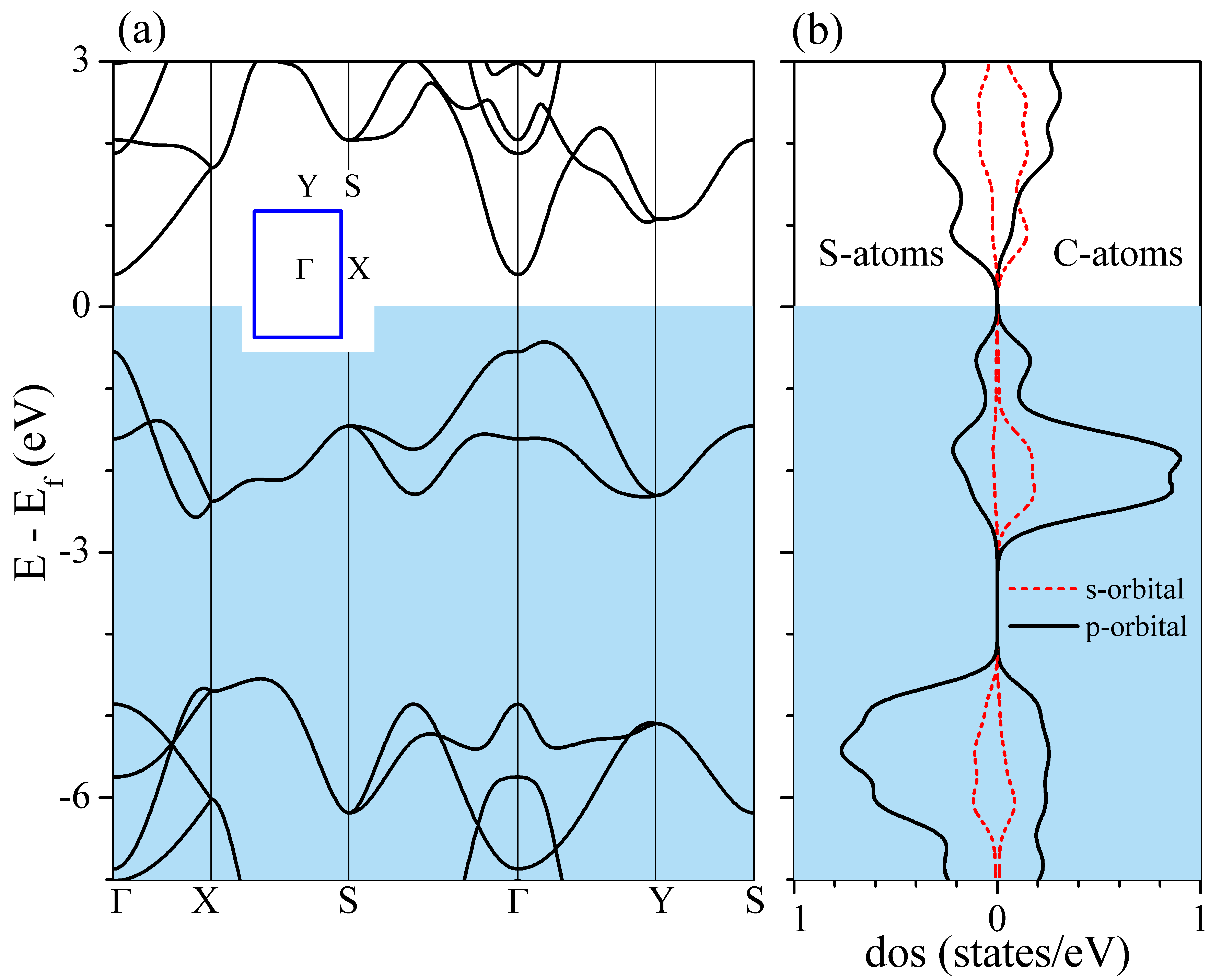} 
\caption{(Color on-line)  Band structure of \textbf{(a)} carbon 
sulfide monolayer and \textbf{(b)} projected density of states. 
Inset shows the Brillouin zone in momentum space and the high-symmetry points. The Fermi level is at $0$ eV. }
\label{Fig:bands}
\end{figure}

The band structure of the carbon sulfide layer in fig. \ref{Fig:bands} reveals a semiconductor with an indirect bandgap of $1.1$ eV, in agreement
with the previous works\cite{alonso2017stable}. 
Slightly higher than the $0.92$ eV bandgap found for phosphorene\cite{fukuoka2015electronic,zhang2016blockage,fei2014enhanced}.
The occupied bands below the Fermi level have a dominant 
contribution of 
$p$-orbitals from carbon atoms. 
Underneath these bands, a particular feature of the CS monolayer 
is the second bandgap located $\approx 3$ eV below the Fermi level.
Followed by bands composed by 
$p$-orbitals from sulfur atoms.

\begin{figure*}[!ht]
\includegraphics[clip,width=0.325\textwidth,angle=0]{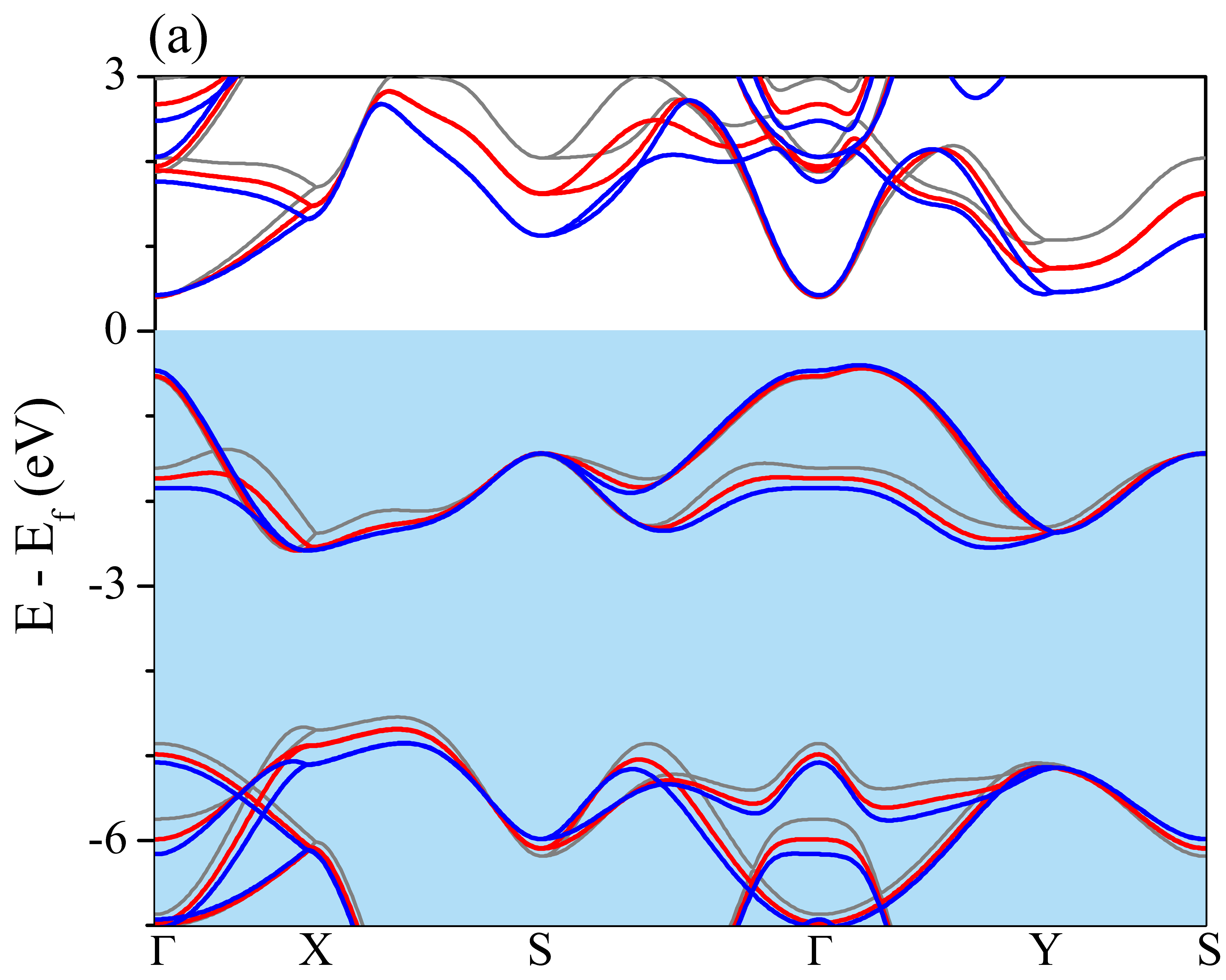} 
\includegraphics[clip,width=0.325\textwidth,angle=0]{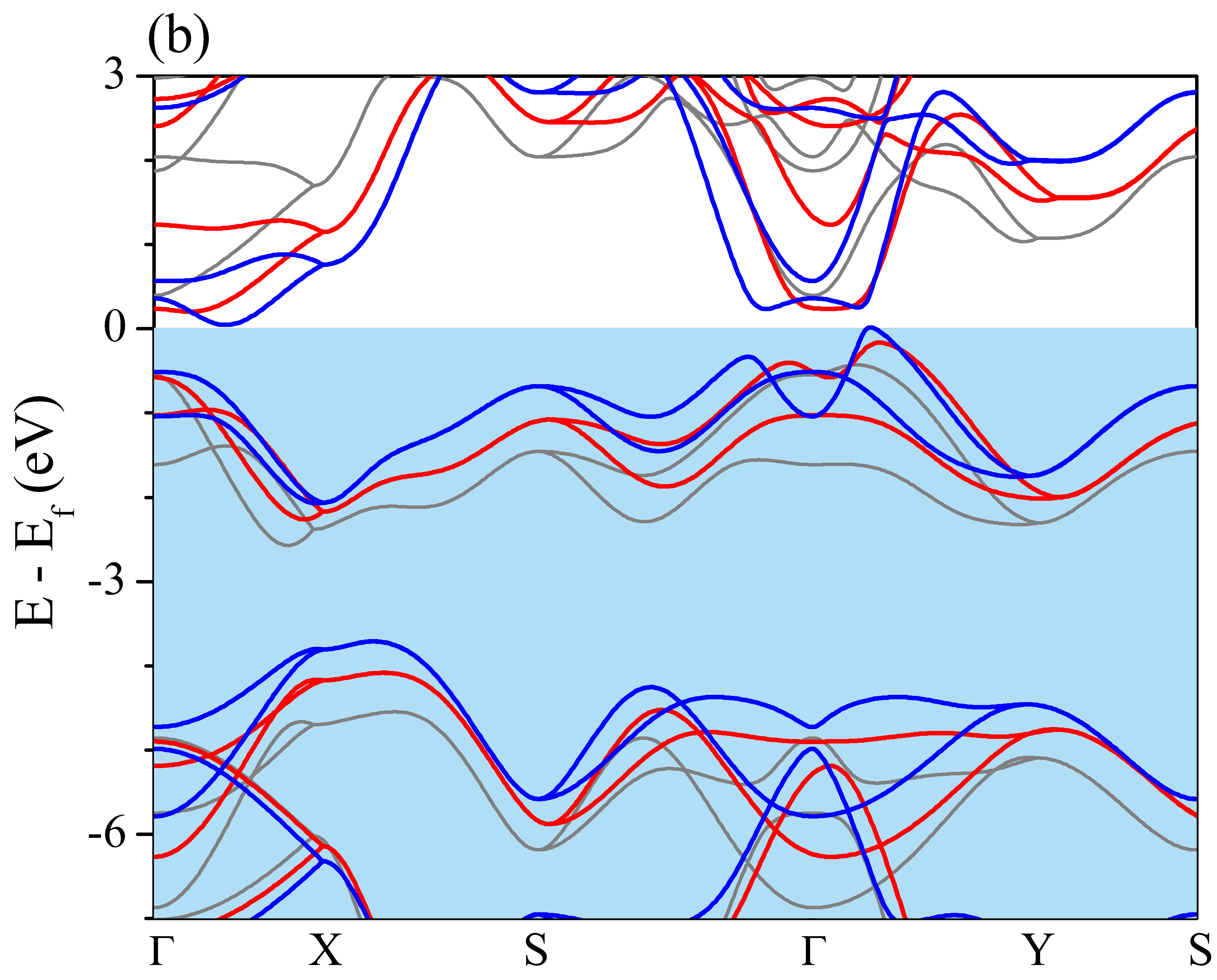}
\includegraphics[clip,width=0.325\textwidth,angle=0]{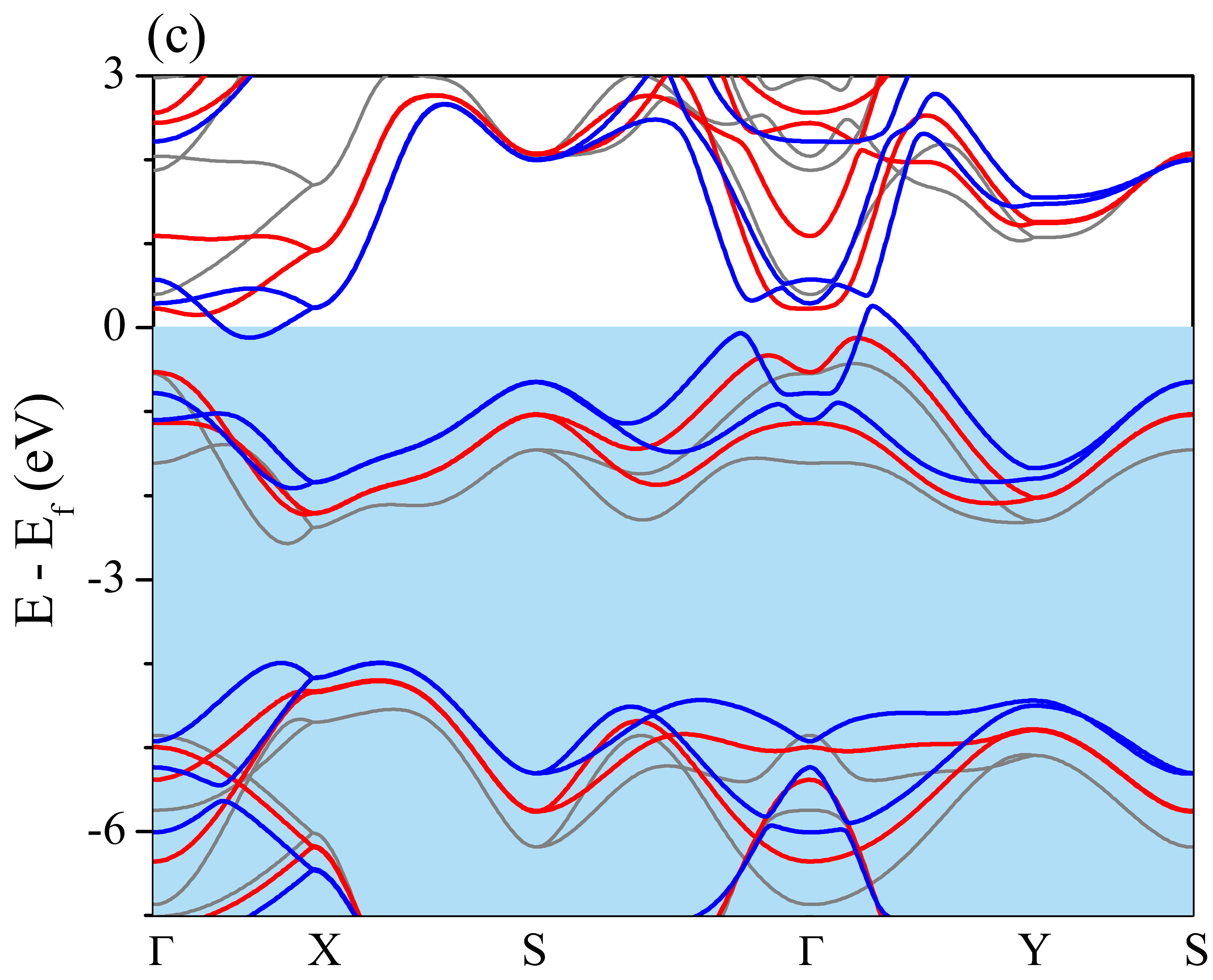}  
\caption{(Color on-line) Band structure of the CS monolayer under 
strain in the armchair-direction \textbf{(a)}, zigzag-direction \textbf{(b)}
and biaxial strain \textbf{(c)}. In each panel, the gray line corresponds 
to the 0\% strain, red is for 5\% strain and blue is for 10\% strain.
The Fermi level is at $0$ eV and the path in k-space is the same of fig. \ref{Fig:bands}. }
\label{Fig:bands_strain}
\end{figure*}

We consider two uniaxial strain cases corresponding to modifications 
of the unit cell in the armchair and zigzag directions, and a biaxial 
case corresponding to a combination of tensile strain in both armchair 
and zigzag directions.
The strain is introduced by changing the size of the lattice vector 
in the selected directions and then the structure is fully relaxed.
Similar to molybdenite (MoS$_{2}$)\cite{scalise2012strain}, 
we find that the bandgap of the the CS monolayer decreases as 
the tensile strain increases. This tendency is the opposite to 
that observed in graphene and phosporene 
layers\cite{si2016strain,nguyen2015enhanced,lv2014enhanced}.

For uniaxial strain along the armchair-direction (parallel to x-axis) 
in fig. \ref{Fig:bands_strain} (a), the band structure is modified
marginally. As the uniaxial tension in armchair-direction increases, 
the bottom of the conduction band at the high 
symmetry $Y$-point of moves down and for a strain of 10\% reaches 
the same level of the $\Gamma$-point.
The bandgap of the CS monolayer is more sensitive to the uniaxial 
strain applied along the zigzag-direction (parallel to y-axis)
as shown in fig. \ref{Fig:bands_strain} (b). 
Contrary to the previous case, the gap closes even for small 
values of strain and the bottom (top) of the conduction (valence) 
band moves away from the high-symmetry points. 
In the fig. \ref{Fig:bands_strain} (c), we can observe that 
the band structure upon biaxial strain (x and y direction simultaneously)
shows  a behavior similar to the uniaxial case applying strain in the 
zigzag direction. 
As the strain increases, the carbon $p$-bands move upward in energy 
producing several anticrossings near the Fermi level.
For strain values of 10\%, the HOMO-LUMO gap has completely disappeared 
and the transition between semiconductor and metal is complete.
Note that, the second gap remains practically constant due to the almost 
rigid movement of the sulfur $p$-bands.

Employing the calculated electronic band structure as input, 
the transport coefficients are calculate using the Boltzmann transport 
theory within the constant relaxation time approximation.  
It is possible to correlate features from band structure with some 
aspects shown by the Seebeck coefficient $S$. Because 
electrons and holes contribute to the transport properties, 
the maximum of the Seebeck coefficient appears in the band 
gap\cite{mahan1989figure,gibbs2015band} and therefore it can be 
controlled with strain engineering. 

In our calculations the maximum Seebeck coefficient around the Fermi 
level $S_{\mathrm{max}}$ at room temperature ($300$ K) for 
the non-strained CS monolayer is $2.62$  mV/K. Under the same conditions, 
for the phosphorene layer we find a $S_{\mathrm{max}} = 2.16$  mV/K, 
in agreement with previous DFT-PBE calculations\cite{fei2014enhanced}.
Note that hereafter we refer to $S_{\mathrm{max}}$ as maximum Seebeck 
coefficient around the Fermi level.
A second peak also appears in the Seebeck coefficient of the CS 
monolayer due to the second gap. We will not discuss it, because 
under normal experimental conditions it would be difficult to reach 
the level of doping necessary to be measured.

\begin{figure}[!ht]
\centering
\includegraphics[clip,width=0.45\textwidth,angle=0]{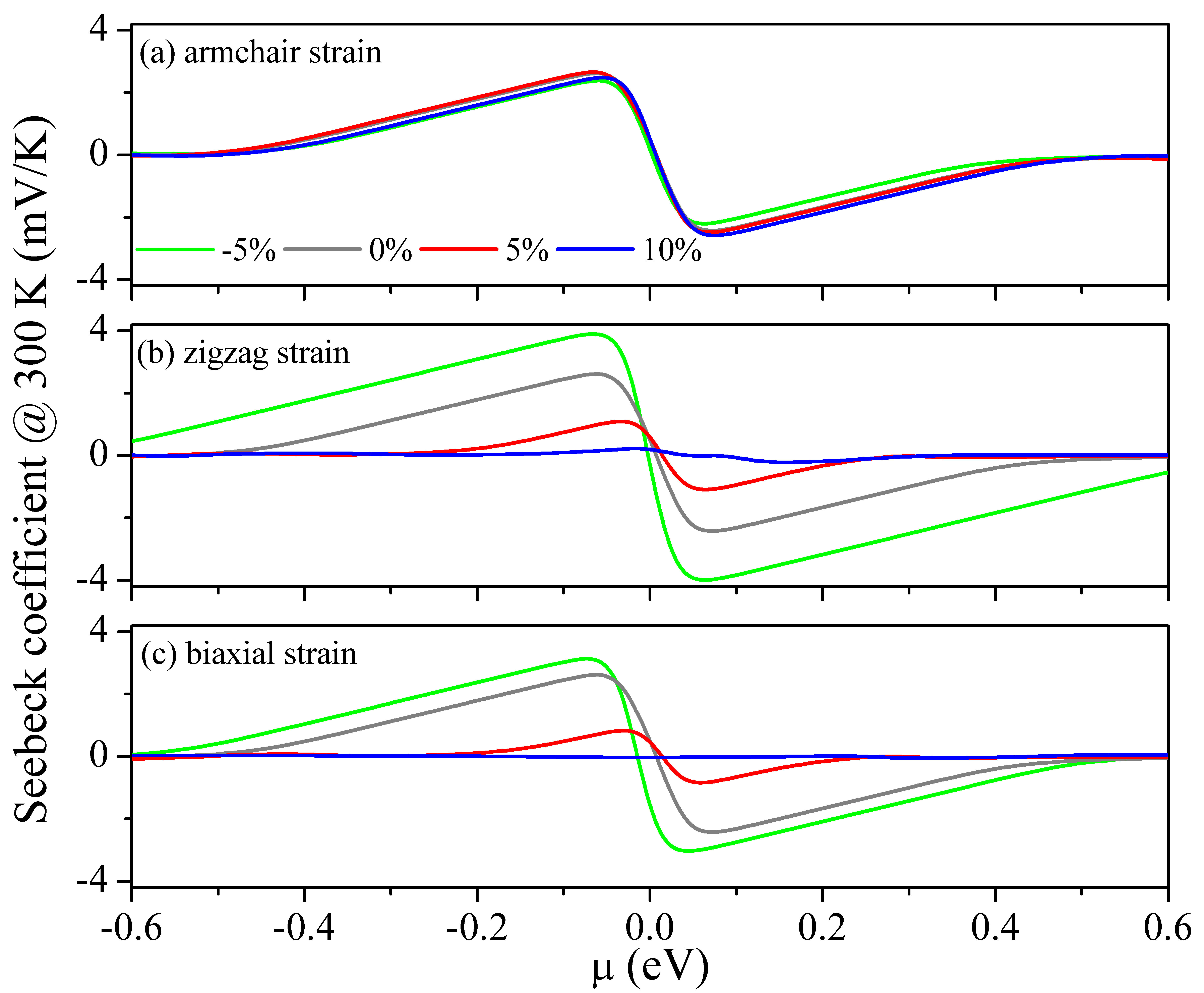} 
\caption{(Color on-line) Room temperature ($300$ K) Seebeck 
coefficient as a 
function of chemical potential for the CS monolayer under uniaxial strain
in armchair-direction \textbf{(a)} and zigzag-direction \textbf{(b)}, 
and biaxial strain in \textbf{(c)}. The different colors are for 
different strain values.
The inset shows the behavior around the $S_{\mathrm{max}}$ for the 
strain applied along armchair-direction (parallel to x).}
\label{Fig:Seebeck}
\end{figure}

In fig. \ref{Fig:Seebeck}, we present the room temperature Seebeck 
coefficient of the CS monolayer as a function of chemical potential. 
The Seebeck coefficient is strongly modified by the chemical potential,
showing that an optimal carrier concentration is crucial for efficient 
thermoelectric performance\cite{fei2014enhanced}. 
The change in the Seebeck coefficient due to the applied strain 
along the armchair-direction is minor, fig. \ref{Fig:Seebeck} (a). 
This result can be expected from the small variations observed in 
the band structure upon uniaxial strain along that direction in 
fig. \ref{Fig:bands_strain} (a).

Transport coefficients are easily controlled by strain in the 
zigzag direction. In fig. \ref{Fig:Seebeck} (b) we can note how 
the behavior of the Seebeck coefficient against the chemical 
potential tends to a constant as the strain increases. 
For negative strain values corresponding to a compression of the system, 
we observe a large increase in the maximum value of the Seebeck 
coefficient. Both behaviors can be explained by the variation of the 
gap with the strain\cite{zhang2014phosphorene}. 

Given the response of the band structure to strain, the 
behavior of the Seebeck coefficient in case of biaxial strain is similar 
to that of the previous case. In fig. \ref{Fig:Seebeck} (c), for 
positive strain values, the Seebeck coefficient follows the same trends
of the uniaxial in zigzag-direction but with slightly lower values.
The difference between both cases appears when considering the 
structural compression, the response of the Seebeck coefficient is 
smaller in the biaxial case.

The maximum Seebeck coefficient decays exponentially with the 
temperature without modifying the behavior described above. 
It is possible to follow the behavior of the maximum Seebeck 
coefficient value $S_{\mathrm{max}}$ for a given temperature.
On the one hand, the biaxial case is more susceptible to
modifications to the positive strain values 
(corresponding to an expansion) where
$S_{\mathrm{max}}$ rapidly tends to zero, the strain in the 
zigzag direction $S_{\mathrm{max}}$ follows the same trends, 
and the strain in the armchair direction only produces marginal 
effects in $S_{\mathrm{max}}$. 
On the other hand, when considering negative strain values 
(corresponding to a compression) in the zigzag direction at room 
temperature and -5\% strain the $S_{\mathrm{max}}$ increases  
to $3.9$ mV/K, the biaxial strain case produces a slight increase 
of $S_{\mathrm{max}}$ to $3.1$ mV/K, and contrary to these two, 
the strain in armchair direction produces a reduction of 
$S_{\mathrm{max}}$ to $2.38$ mV/K. As a reference, at $300$ K for 
the non-strained case we find a $S_{\mathrm{max}}=2.62$ mV/K. 

As previously noted, the electronic and thermoelectronic responses of 
the CS monolayer are contrary to that observed in graphene and 
phosporene layers\cite{nguyen2015enhanced,lv2014enhanced}.
Taking advantage of the fact that the Seebeck coefficient is low 
for conductors and the electronic conductance tends to be low for 
insulators\cite{chico2017spin}. Devices could be designed taking
advantage of inhomogeneous response upon strain using 
heterojunctions or controlled stacking to obtain a 
strain-dependent thermal and electronic 
responses\cite{lado2013quantum,shim2016phosphorene,rosales2013transport,dai2014bilayer,gonzalez2010electronic,deng2014black}. 

\section{Final Remarks}
In conclusion, we have investigated the strain response on the 
electronic and thermoelectric properties of carbon sulfide 
monolayer 
based on the PBE-DFT calculations combined with the semiclassical 
Boltzmann theory.
We found a highly anisotropic electronic and thermoelectonic 
response upon strain. When the strain is applied in the armchair 
direction, the bandgap and Seebeck coefficient remain almost unchanged. 
In contrast, when strain is applied in the zigzag direction the Seebeck
coefficient is easily modulated, going from a finite value to zero with 
relatively small strain values. 
By tracking the evolution of the Seebeck coefficient as a function 
of external strain we can follow the change in bandgap induced by 
the strain. Our results suggest possible applications as 
sensors or active component taking advantage of the real-time 
bandgap modulation.

\section*{Acknowledgements}
The author gratefully acknowledges the critical reading of the manuscript and the suggestions made by L. Chico and S. Sadewasser.



\end{document}